# QUALITY OF SERVICE-AWARE SECURITY FRAMEWORK FOR MOBILE AD HOC NETWORKS USING OPTIMIZED LINK STATE ROUTING PROTOCOL

Thulani Phakathi, Francis Lugayizi and Michael Esiefarienrhe

Department of Computer Science,
North-West University, Mafikeng, South Africa

## ABSTRACT

*All networks must provide an acceptable and desirable level of Quality of Service (QoS) to ensure that applications are well supported. This becomes a challenge when it comes to Mobile ad-hoc networks (MANETs). This paper presents a security framework that is QoS-aware in MANETs using a network protocol called Optimized Link State Routing Protocol (OLSR). Security & QoS targets may not necessarily be similar but this framework seeks to bridge the gap for the provision of an optimal functioning MANET. This paper presents the various security challenges, attacks, and goals in MANETs and the existing architectures or mechanisms used to combat security attacks. Additionally, this framework includes a security keying system to ascertain QoS. The keying system is linked to the basic configuration of the protocol OLSR through its Multi-point Relays (MPRs) functionality. The proposed framework is one that optimizes the use of network resources and time.*

## 1. INTRODUCTION

A MANET is an autonomous type of system which has separately connected sets of self-configuring nodes that may be activated by putting to use various techniques e.g. Bluetooth or WLAN. MANET is autonomous in behaviour because each node is regarded as its host and router at the same time [1]. They rely on direct communication and multi-hops for communication between distant nodes within the network making them scalable and robust. These advantages make them more flexible to accommodate many nodes, decentralize administration and their setup can be placed anywhere and at any time [2]. Contrary to other Wireless systems, MANETs do not have a central authority that monitors the forwarding of traffic. MANET systems consist of an infrastructure-less system of associated nodes connect through wireless links [1]. Nodes move arbitrarily or rather in a random faction [3]. Security in MANETs is crucial from the node level to the network level. According to [4], because of the dynamic nature of MANETs, trust can be used as a measure for nodes that want to provide an acceptable level of trust in that relationship among themselves. Security in a MANET is way too challenging than in traditional network environments infused with a central controller because of the dynamic topological nature and characteristics of MANETs. MANETs are primarily used in the army and security-based applications e.g. covert missions, emergency, and rescue missions [8] [6] [9]. The availability of network resources, integrity, and confidentiality depends on the security mechanisms that are put in place. The design of MANETs makes them vulnerable to security attacks. The vulnerabilities come through non-secure boundaries and compromised nodes although many other factors contribute to the weakening nature of MANETs. The best approach to mitigating attacks is prevention and avoidance algorithms, not security mechanisms that remove attacks as these tend to require more resources. Network resources in MANETs must be optimally used to achieve Quality of Service.  The needs for the provision of QoS are increasing with applications that involve voice and video and it is most appropriate to support these through the implementation of ad hoc network environments. QoS was first brought to attention in 1994 as a phenomenon that has the overall requirements of a network connection, as well as response time during service times, network detriments such as echo, interrupts, signal to noise ratio and also loudness levels. QoS is generally the network's assurance to ascertain a specific level of execution to a data transmission [10]. To achieve QoS, the concept of routing is important. Routing is regarded as the act of steering information from a source node to a terminal node in the network. [10]. One intermediate node within the network is experienced during the movement of information. The most important aspect is achieving good QoS. It is impossible to say a characteristic like this one can be completely run over. It is possible to achieve a greater QoS to such an extent that its dynamic nature would not be such a limitation thus a robust and efficient security framework is needed. The framework would guard against malicious activity in the network amongst nodes. This work seeks to close that gap by building a framework that not only looks at the security but also the Quality of Service in video streaming applications over MANETs.

This work is arranged as follows: Section 2 discusses the various routing protocols (RPs) in MANETs, Section 3 is on trust in MANETs and Section 4 presents security frameworks studies. Section 5 presents Typical security attacks in MANETs. Section 6 gives a highlight of related works while Section 7 presents the proposed framework in detail.

## 2. OPTIMIZED LINK STATE ROUTING FEATURES

### 2.1. OLSR Protocol

OLSR is simply the optimization of traditional link-state protocol for MANETs. It falls under the category of proactive routing protocols. With the OLSR protocol, every network node chooses a neighbouring node-set, commonly termed as the multipoint relays (MPR), which rebroadcasts the packets that were initially transmitted. To this end, neighbouring nodes that are not found in the MPR set have the instruction to only read and process the packets [14]. According to Saravanan and Vijayakuma [18], OLSR keeps tracks in the pathfinding table to provide a route if necessitated. MPRs are primarily responsible for declaring and forwarding link-state information, forwarding and controlling traffic, providing effective mechanisms for broadcasting control traffic by minimizing the frequency of required transmissions [19]. OLSR utilizes two types of control messages: Hello and Topology Control (TC). Hello messages are used to the information concerning the link status and the host's neighbours [20].

### 2.2. OLSR Architectural Design

OLSR has a cross-layered design just like that of the OSI (Open Systems Interconnection) model in networks. From a designer's perspective, there are two relative choices in the design process of the protocol. The first option is to design the protocol by the rules of the reference architecture and that is the higher layer being able to access services provided by the lower layer with no consideration of how such service is made available. Secondly, the routing protocols can be developed in violation of the original architecture and that is by allowing cross-layer communication between layers. This deliberate violation is called the cross-layer design.

## 3. SECURITY GOALS IN MANETs

The goal of every system is to achieve excellent quality of service and adherence to security targets to protect client data or information [24].

### 3.1. Availability

Availability is one security target for every system. An authorized user will request us of the node and in an operable state. The node is therefore supposed to provide its services by its design. Attacks may seek to disrupt the node's operation and also use up some of the node's resources but the node must be able to survive those attacks and be available when requested

### 3.2. Confidentiality

This security feature ensures the unavailability of certain features to unauthorized entities or users. The information is restricted to only authorized personnel. A message that an as the source will only be decrypted at the destination node. Many cryptographic attacks may try to reveal the message contents. An ideal system will be able to protect the contents of such information from unauthorized users.

### 3.3. Non-repudiation

This feature ensures that the source node will not be able to interfere with an occurred action like deny the authenticity of a message sent. It also facilitates the detection of malicious nodes. Many of the existing algorithms are based on reputation and trust e.g. CONFIDANT.

### 3.4. Authentication

Authentication ensures user validation and avoiding impersonation. The malicious node could impersonate a legitimate node by using the node's MAC address or even an IP address to obtain authentication and also launch its attack at a higher level.

### 3.5. Integrity

Integrity is a security feature that ensures that the original contents of the data are maintained and not altered in any way. An effort to intercept the data being transmitted, either by human beings or malicious nodes is an act against the trustworthiness of the network. Dropping attacks are usually launched by a malicious node but the node is compelled to cooperate in the system.

### 3.6. Anonymity

This feature is for privatizing the true identity of a node to ensure privacy and confidentiality. In most cases, the source of packets is kept private.

## 4. VULNERABILITIES IN MANETs

MANETs are prone to various [13] security vulnerabilities that pose to gain unauthorized access to the user's data. Vulnerabilities of a system may be termed as weaknesses possessed by a system. MANETs are more vulnerable because they rely on wireless technology, unlike traditional wired networks. Attacks in MANETs can be categorically put in two; namely active and passive attacks. The severity of these attacks differ. Wireless networks are most vulnerable to all sorts of attacks internally and externally as compared to [28] traditional networks (wired networks) due to limited physical security, scalability, mobile nodes, dynamic topology, lack of centralized management, and threats emanating from compromised nodes. The vulnerability of the network highlights a weakness in the security architecture or system. MANETs operate in very dynamic environments. Security is very important in MANETs even though the environments' hostility makes it difficult to achieve most security properties as proposed by many authors [24].

### 4.1. Central Controller

The lack of a centralized controller that could act as a monitoring-server is one vulnerability that comes with MANETs. This makes it complex in terms of security provision against attacks as in most instances the network environment is huge and highly dynamic.

### 4.2. Dynamic Topology

As mentioned earlier, the network environment in MANETs is dynamic. This may, in turn, affect the trust relationship among nodes. Malicious nodes that may be compromised within the network are also difficult to spot as the nodes are mobile.

### 4.3. Power Limits

Mobile nodes rely solely on battery power and such may pose so many problems. A node may behave maliciously within the network and could be suspected of being an internal attacker but only to discover that it behaves selfishly because of limited power supply.

### 4.4. Resource Availability

Resource constraints are the primary reason why some services are not utilized in MANETs. For example, secure communication is needed but most often it is difficult to provide it because of the dynamic environment. Ad hoc security mechanisms and architectures are needed to prevent attacks from flooding the network.

## 5. SECURITY ATTACKS IN MANETS

Attacks in MANETs can be categorically put in two; namely active and passive attacks. The severity of these attacks differ. Wireless networks are most vulnerable to all sorts of attacks internally and externally as compared to [28] traditional networks (wired networks) due to limited physical security, scalability, mobile nodes, dynamic topology, lack of centralized management, and threats emanating from compromised nodes. The vulnerability of the network highlights a weakness in the security architecture or system.

### 5.1. Active Attacks

These are known to disrupt the normal operation of a network [21]. The attacker actively alters the network's normal operation. The attacker acts as one of the stations in the network. In this way, it able to exploit any other node and uses it to its advantage. It can feed nodes fake packets or even denial of service (DoS). The active attacker can:

- Fabricate messages
- Replay packets
- Modify packets
- Drop packets
- Node Impersonation
- Insert infected code

## 5.2. Passive Attacks

Passive attacks are characterized by their inability to [19] actively participate in causing harm to the network. The attacker monitors the network to attain information. They do that so that they may get node information, for example, how nodes are communicating and their geographical location within the network. They do not just attack the network. At first, they acquire enough information before launching an attack. Once they acquire information, they easily hijack it and launch an attack. They can decrypt weakly encrypted data, acquire passwords, private and public keys, monitor communication routes, and message flow among entities [29]. It may be hard for the user to identify a passive attack as it does not necessarily alter anything regarding user data or traffic.

## 6. RELATED WORKS ON SECURITY FRAMEWORKS IN MANETs

Hurley-Smith et al. [31] proposed a security protocol called Security Using Pre-Existing Routing for Mobile Ad hoc Networks (SUPERMAN). The initial protocol design was to solve issues like the authentication of nodes, secure network access control, and secure network communication through existing routing protocols. SUPERMAN was designed to bring together communication security and routing at the network layer. Their security protocol is unique from others in the sense that it combines routing and communication security at the network layer. This is in contrast to existing approaches that may require additional protocols to protect the network. SUPERMAN was created to give security to all data communicated over a mobile ad-hoc network. It may not apply to other networks. In a nutshell, it provides protection and efficiency. One efficient method it employs is that it protects application data and routing, ensuring that the network provides trustworthy, confidential communication, and reliable to all true nodes [31].
Kaur et al. analyzed [33] security threats MANETs. Their security objective called CBDS was successfully carried out on Blackhole and Grey hole attacks before and their trial was proven successful in the case of Sleep deprivation and denial of service attacks. Their simulation results have showed increased detection for CBDS and an enhanced response [35]. The limitation within the framework was its implementation of two attacks and no proof is given out in terms of its validity towards other active and passive attacks.

Monica et al [35] in their work analyzed, simulated and three different attacks based on many parameters. These attacks were Blackhole, Denial of service (DoS, and wormhole. The comparison was made for their throughput, End to End Delay, and Packet Delivery Ratio (PDR)
Authors in [5] proposed a MANET trust model that uses a combination of direct, indirect, and mutual trust values among network nodes to reflect the behavior of sensor nodes. The aim was to test out the effectiveness of a secured node can be routed within the network. QoS metrics were used to evaluate the trust level. This provided an accurate recommendation for packet forwarding and thus reducing the rate of dropped packets. A performance evaluation was conducted and the trust model achieved reduced packet loss, reduced energy consumption, and an increased network throughput despite having malicious nodes in the network. This meant that that proposed algorithm improved the overall network performance as compared to an existing single-trust based model. The proposed model filtered out malicious nodes.

Sahu et. al [7] proposed a cross-layer security framework that prevents malicious attacks in the network. The proposed framework used throughput, end to end delay, jitter, and packet drop ratio as QoS metrics where-in different scenarios were evaluated. Their framework included a design and implementation of a Neighbour Node Surveillance Real-Time MAC(NNSRT-MAC) protocol at MAC Layer. Key contributions include the design and implementation of a QoS framework that doesn't use a complex algorithm to prevent over-reservation, QoS degradation, flooding attack, state table starvation. The results gave better results in terms of the QoS metrics in both malicious-free and malicious scenarios.

Madhavan et. al [11] proposed an algorithm called GA-ACO (Genetic Algorithm-Any Colony Optimization) to optimize QoS by using a secure agent-based multicast routing scheme to optimize parameters by combining GA and ACO techniques. This hybrid technique outperformed existing protocols like AODV and OLSR.

Authors in [12] proposed an efficient multi-hop and relay-based communication framework. Using Brodatz Texture database, CT scan images, and Brain MRI scan images as input. The algorithm was designed to operate 3 stages. The first stage involved selected regions using the spatial candidate region detection. The second stage involved applying average entropy feature space for the detection of the cluster centre. The final stage involved spatial density-based clustering of images carried out by tracking down dense regions. This method produced better clustering results and PSNR rates. The improvement of QoS was based on Random Repeat Trust Computational Approach using direct and indirect trust. The proposed framework showed more than 30% effectiveness as compared to the existing system.

Tygi et. al [17] proposed a Proposed Local Adjustment AODV in high mobility environments with scarce network resources and ruptured explored routes The algorithm controls the flooding of control packets and its maintenance during transmission with maximum adjustment locally when a node responsible of forwarding packets is out of range in terms of transmission. Metrics used were control overhead, packet delivery ratio, and energy consumption. The proposed algorithm outperforms AODV in terms of the QoS metrics evaluated.

Authors in [25] investigated the routing protocol ZRP by improving an existing algorithm called zone-based routing with parallel collision guided broadcasting protocol The network's topology is controlled using an estimate of the node's energy dropout rate. The energy efficiency is measured enhanced to find an optimal QoS routing path and reduced overhead. The proposed protocol gave improved results in terms of performance as compared to other experimented protocols

The architecture presented in this work is a conceptual model of the proposed framework. The framework's cross-layered design is strongly in-line with OLSR's architectural design in that the higher layer can access services provided by the lower layer with no consideration of how the service is made available. The overall framework utilizes the TCP/IP model to fully articulate each stage. The layers exhibited are the data link layer, application layer, physical layer, network layer, and the transport layer. model. The layers exhibited are the data link layer, application layer, physical layer, network layer, and the transport layer.

## 7. PROPOSED QOS FRAMEWORK & ITS OPERATIONS

This section presents a conceptual (fig.1) and a high-level overview of the proposed security framework and the different components or technologies associated with it. It also presents assumptions that we made in connection with the framework.

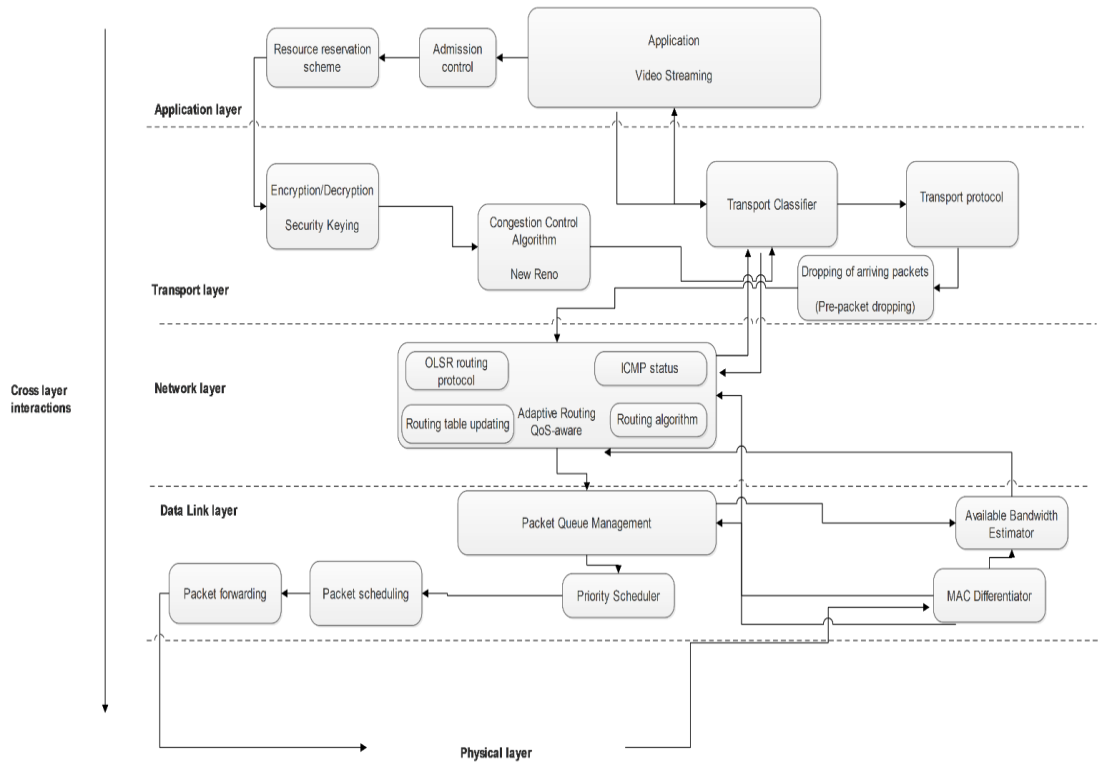

Figure 1.  QoS-aware security framework

## 7.1. Application Layer

The application layer is mainly responsible for generating relevant traffic i.e. CBR, video, voice, email, and HTTP. To fully ascertain QoS, a resource reservation scheme is implemented. A resource reservation scheme will ensure that QoS for high priority sessions is guaranteed and that sufficient bandwidth is administered throughout the transmission phase to fulfill the fundamental requirements of QoS. The assigned resource reservation scheme will guarantee QoS performance as it decreases the chances of high priority session collisions while using the bandwidth. Security threats like malicious nodes, worms, and viruses ar. The end goal of the framework is the assurance of QoS within the network. This can be achieved through:

- QoS Medium Access Control (MAC) scheduling
- Admission Control
- QoS-aware routing
- Traffic policing

These four mechanisms are covered throughout the framework at different layers in terms of implementation. The proposed QoS-aware routing protocol solution will be based on:

- OLSR protocol and QoS
- OLSR protocol and MAC protocol

## 7.2. Transport Layer

At the transport layer, there are different activities involving the movement of packets from the application to the network layer. It focuses on end to end communication during data encryption. This can be done using the Transport Control Protocol (TCP) or User Datagram Protocol (UDP) protocol.

### 7.2.1. Congestion Control Algorithm

This is used to control the congestion within the network. In our work, we implement New Reno, an algorithm that improves retransmission during the quick-recovery phase of TCP Reno.

### 7.2.2. Transport protocol

 A transport protocol has the prime responsibility of establishing and facilitating the movement of data from one node to the other. In this work, the TCP protocol is used because of the streaming data traffic.

### 7.2.3. Encryption

This is an effective way of achieving data security. A secret key will be used to have access to an encrypted file. This is called Decrypting. This work uses AES (Advanced Encryption Standard) algorithm.

## 7.3. Network Layer

The network layer is more central to the realization of a fully functional system whereby there is bi-directional communication between the transport and network layer moreover the network layer and the data link layer. This includes:

### 7.3.1. Adaptive routing (QoS-aware)

A process of determining the most efficient path in which a data packet can use in a network to reach its destination

### 7.3.2. Routing protocol

It determines how MANET nodes should communicate with each other by round-robin fashion that enables them in selecting optimal routes between any two nodes within the network.

### 7.3.3. ICMP status

This relays messages about the status of our IP address e.g. Destination Unreachable, Time exceeded and Trace Route

### 7.3.4. Packet scheduler and buffer

This contains the actual memory that is used to store packets. Additionally, the scheduler automatically builds a protective front (firewall) against hostile nodes and thus protecting network resources from saturation.

### 7.3.5. Packet classifier

This process strategically categorizes packets into flows. It contains a set of rules categorizing packets according to their header fields

### 7.3.6. Routing algorithm

This is a set of stepwise operations implemented to direct internet traffic more efficiently. It mathematically determines the best path to take during routing in the MANET.

### 7.3.7. Link state table domain

This a table that contains link information about all known MANET nodes exercising routing functionalities.

### 7.3.8. QoS scheme

These are mechanisms utilized for the attainment of an acceptable level of QoS in the network.

### 7.3.9. Admission control

This is a very important component of the proposed framework. It is key in the provisioning of QoS in the MANET because it determines the fair provisioning of network resources and the extent at which they are utilized. and if QoS characteristics are delivered. Admission control can be considered as a validation process where the checking is done before the establishment of a connection to calculate the sufficiency of network resources for a proposed connection.

## 7.4.  Data Link Layer

The IEEE 802.11 standard is utilized at the data link layer. The MAC plays a huge role in linking the network and physical layer. The MAC address plays a central role in handling queues during routing and bandwidth estimation during cross-layer interactions between the data link layer and the physical layer. The responsibility of the link layer in this framework is towards the MAC Access control and it performs the following activities;

- Packet scheduling and forwarding
- Priority classification
- Packet Queueing
- Bandwidth Estimation

## 7.5.  Framework Operations

The conceptual view of the framework presented above demonstrated some of the key stages in the actual prototype. The prototype in Fig 4.1 follows the same design mechanism like the one in Fig 3.6. The TCP/IP model is at the core of the design of the framework. At the upper level is the application layer which contains video traffic generation, admission control, and a resource reservation scheme which links up with the keying system at the transport layer.

The keying system is responsible for security encryption and decryption to prevent attackers from having access to information passed on from the source to the destination. The utilized encryption method is an improved AES standard as shown in Fig 3.7 of chapter 3. A proper data handover architecture was designed to fully support MANET environments. The alternative route in the security structure was designed to maintain data integrity and prevent passive attacks which may phish for information in the channel and observe activity within the channel. The next step would be the congestion control algorithm which in this instance is New Reno, an algorithm that improves retransmission and helps share. The transport classifier then takes over to the transport protocol which in this instance is Real-Time Messaging Protocol. This protocol is chosen over UDP and TCP based on its excellent delivery in video streaming traffic although traditional networks would opt for TCP as it guarantees packet delivery other than UDP which does not allow retransmission. Cross-layer interactions between layers (application and transport) continue because of the transport classifier which links up with the application.

The network layer is primarily responsible for QoS aware routing and that involves a technique called adaptive routing. Adaptive routing determines the best/optimal path a data packet should follow in the network to reach its preferred destination. This is achieved by using the shortest path algorithm which takes the data packet to destination with minimal congestion and thus efficiently using the network resources. This algorithm allows nodes to calculate routes in given network topology and thus saving time and minimizes overhead size. Adaptive routing improves network performance as routes adjust automatically in response to dynamic network topology. The nodes exchange route information and updates during adaptive routing.

This is necessary to fully adhere to QoS requirements. The routing protocol, OLSR, facilitates routing in the network layer. It is the highest decision making entity that is responsible for directing all packets. The routing table is constantly updated as the node moves randomly within the network. Most of the operations that happen in our framework depend on the network level. The ICMP status is for monitoring the IP connection of the network.

The data link layer provides more of a link address (MAC) to associate with the IP address shown at the ICMP status. The framework has an available bandwidth estimator. These addresses work to ensure that packets are properly scheduled, forwarded, and allocated appropriate resources (bandwidth) and that each node's unique identifier (MAC address) is linked with an IP address. There are so many cross-layer interactions within this layer as the physical layer is also involved. When packets are forwarded to the physical layer then the process of moving from the physical to application layer begins. The packets will move from the MAC differentiator to the QoS aware mechanism, link up with the transport classifier, and lastly the application, where it all started.

### 7.5.1. Security Keying

There are several threats to the security of a MANET but again resource allocation plays a huge role build-up to a practical security structure. The Key management approach, AES, in this work is implemented because it prioritizes primary data protection and security. In traditional networks, Watchdogs and controllers are used to impose a well-functioning structure but involving such tools would surely drift away from the MANET architecture as MANETs have no central authority but every node acts out as its independent router and regulator. The proposed framework aims at achieving an acceptable level of QoS by securing the network from active and passive attacks. Figure 2 shows a keyed source node sending packets towards its intended destination.

The node broadcasts information to the network and as expected would have alternative route links (marked as Atl. Link in the figure). The information. Alt. Link 4 was able to decode the information from the source and will now be used as a backup in an event that the received information was incomplete or compromised. It then sends the information to the terminal node. The terminal node will then compare the two and compare the signal received from both the original link and the alternative link. This approach not only saves network resources but fully subscribes to the operation of MANETs in terms of the broadcast mechanisms used. There is no need for additional network devices to fully carry security. The primary focus would now be to provide good QoS in terms of end to end delay, packet loss, and throughput. The keying system depends on the functionality of the routing protocol called Multi-Point Relays (MPRs).

Every network workstation chooses a neighbouring node-set known as the MPR. OLSR is considered to be a proactive routing protocol for mobile ad-hoc networks. It uses a link-state algorithm for routing and thus making routes available immediately whenever it needs them. OLSR uses multipoint relays (MPRs) to minimize the overhead from broadcasting control messages. Through MPRs, the protocol can significantly reduce the number of retransmissions that are needed to broadcast a message to all network workstations. OLSR provides shortest path routes through the flooding of a partial link state. OLSR periodically maintains destination routes in the network. Another advantage of OLSR is that it regulates the reactivity of topological variations by decreasing the highest time interval for periodic control message transmission. This means more optimization can be achieved in denser network environments. This result is far better as compared to the traditional link-state algorithm.

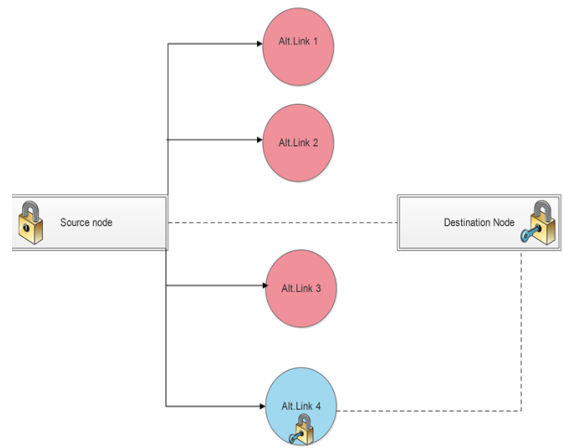

Figure 2. Security keying system

The security keying system used in the framework is the AES Standard for both encryption and decryption. AES [1]is chosen on the basis that it consumes fewer resources (e.g. battery power), fast, and most effective algorithms.

AES is sometimes referred to as the Rijndael block cipher operating on matrix blocks with 8-bit entries of size:

$$4 \times N_b \qquad\qquad (1)$$

Whereby $4 \leq N_b \leq 8$ is the block length.$N_b$ represents the number of bits.

Encryption scrambles the message and outputs it as unrecognizable data. Decryption takes the encrypted data which would be in the form of unrecognizable data and adds a source key to output the original message. The AES is proposed because of its less resource constraint, lightweight, and less computationally demanding features when it comes to routing. The AES is made up of a couple of initialization steps. The first step is the key expansion where the key is broken down to multiple subkeys and the other step is the initial round which involves substitution and transposition. The keys can be broken down into the following:

Table 1 AES algorithm processing properties [2]

| Bits | Cycles | Rating |
|------|--------|--------|
| 128-bits | 10 | Fastest |
| 192-bits | 12 | Slower |
| 256-bits | 14 | Slowest but most secure |

In this work, the 128-bit version of key management is implemented. A series of rounds are performed using the multiple sub-keys made during the expansion phase. The number of times the rounds are made depends on the size of the key we select as highlighted in Table 1. As part of the contribution of this work, we introduced an intermediate trusted node between the source and the destination. This unique node can relay the same message packets issued from the source node to the destination. It provides an alternate route to secure the integrity of the message. This is done to ensure that at the destination, all packets are delivered. In an event whereby Node A, for example, cannot get all the packets to Node D, then Node X acts as surety for complete packet delivery depending on the routing table as repetitive packets are discarded. This would mean that Node X is equipped with encrypted relay capabilities to Node D.

Node X acts as a Multi-Point Relay (MPR) node. MPR's are trusted nodes within the network to relay routing information to the intended destination. This would practically mean that Node A selects Node X as an MPR then retransmits control packets from Node A. In the network, each transmitting node could have one or more of these MPR nodes. These are nodes that are selected by their 1-hop neighbours to retransmit all the broadcast messages it receives from that particular node provided that message is not a duplicate and that the message has a "Time to Live" field greater than one. Routes are selected by MPRs to avoid data packet transfer problems over uni-directional links. Each node will select its MPR set by using its 1-hop neighbours.

A group of MPR nodes is called the MPR set. The set is chosen such that it covers all symmetric 2-hop nodes and a coverage strictly confined within the radio range. An MPR set of Node X is denoted as MPR (N). The other nodes within the network not selected as MPR process control packets as it is a broadcast environment but do not forward the packets. If Node A has selected Node X and Y as its MPRs then it is safe to say:

$$\text{Node A: MPR(A)} = \{X,Y\} \qquad\qquad (2)$$

Where X and Y represent Node X and Node Y, MPR (A) is the set of MPR nodes belonging to Node A.

The MPR nodes are select based on a neighbour basis to the transmitting node. Each transmitting node uses HELLO messages to determine its MPR set. These HELLO messages are periodically broadcasted to one-hop neighbours and not forwarded. Through the neighbour list in the HELLO messages received, nodes can determine 2-hop neighbours and an MPR set. This MPR set is assigned a sequence number and the sequence number is incremented each time a new set is calculated. An MPR set is re-calculated when there is a change in 1-hop or 2-hop neighbourhood detected. MPR nodes are the only ones allowed to generate and propagate Topology Control (TC) messages. The advertisement before sending TC messages is not sent to all links in the network. MPRs minimizes the control traffic overhead of OLSR through retransmission of control messages. The technique significantly reduces the rate of transmissions needed to flood a message to all network nodes. The introduction of MPRs is to minimize the overhead of flooding messages and reducing redundant retransmissions in the network.

The encryption and decryption [2] methods may appear similar but in essence function differently and need to be separated. Rounds are several repeated AES repeated at a set number of times. Encryption has the following steps from round 0 till 9:

- Byte substitution
- Shift rows
- Mix columns
- Add Round key

In pseudo C notation, the above is derived as:

Round(State, RoundKey)
{
ByteSub(State);
ShiftRow(State);
MixColumn(State);
AddRoundKey (State, RoundKey);
}

For the last round execution (round set 10) for AES Encryption is presented in the following order as shown below:

- Sub byte
- Shift Row
- Add Round Key

In pseudo C notation, we can represent it as:

FinalRound(State,RoundKey)
{
ByteSub(State) ;
ShiftRow(State) ;
AddRoundKey(State,RoundKey);
}

Decryption has the following steps for the rounds 11 till 15:

- Add Round key
- Mix Columns
- Shift columns
- Byte substitution

For the last round of operation AES decryption has the following steps:

- Inverse Shift Rows
- Inverse Sub bytes
- Add Round Key

When reducing the number of rounds performed, this may reduce power consumption but would harm the security of the protocol by making it less secure. The 10 rounds of key expansion imposed on this work are done to strengthen the security of the protocol. Naturally, seven rounds or more can be considered fairly secure and energy-efficient.

### 7.5.2. Congestion Control Algorithm

In the framework, the presence of a congestion control algorithm is clearly outlined. Traffic load is one of the most important factors to be considered when addressing QoS. This is because the network resources and scalability thereof are tested within the transport and network layers respectively. When there are too many packets within a network, this may cause packet delay and loss. Packet delay and loss may affect the performance of the system and such a situation is called congestion. It is for that reason the cross-layer interactions are important among the transport and network layer as both layer share in the responsibility of safeguarding traffic load. When there is congestion, it would mean that the available network resources are limited/less than the load. It is the responsibility of the network to resolve congestion. For efficient operation, it is better to reduce load as highlighted in the framework (figure 4.1)

$$Efficiency = \text{Pre-packet dropping of arriving packets} \qquad (3)$$

Another possible solution towards congestion control would be to increase resources which is the job of the Admission control component of the framework. Admission Control is done before a connection is established so it is virtually impossible to increase resources in the middle of transmission as such a validation process is performed before transmission.

In this work, a congestion control algorithm called New Reno is used. It is an algorithm derived from the algorithm called Reno, which was proposed because of the inefficiency of Tahoe. Old Tahoe is the combination of the slow start and congestion avoidance algorithm. The later version of Old Tahoe is called Tahoe. Tahoe is an algorithm that works on duplicate ACK whereby retransmission happens without waiting for a timeout. During packet dropping, Reno enters fast recovery multiple times and thus decreasing the congestion by half. In scenarios where multiple packets are being dropped Reno does not, however, increase throughput. New Reno, the modified version, uses TCP to store a sequence of number that belongs to the highest data packet. New Reno implements fast recovery in the case of three duplicate acknowledgments and improves retransmission during the fast recovery phase of TCP Reno. When the partial ACK arrives, the congestion window is severely reduced by the amount of acknowledged data after the retransmitted packet. This acknowledged data is then called new data as shown in equation 3:

$$cwnd = cwnd - \ + SMSS \qquad (4)$$

Where, Cwnd being the congestion window, is the new data and SMSS being the Sender Maximum Segment Size.

### 7.5.3. QoS-aware adaptive routing

The routing protocol, OLSR, facilitates all routing in the network. OLSR works using a link-state algorithm that constantly works with the routing table. The routing table is flexible to adaptive routing as the topology is dynamic and it needs constant updating. The packets will use an optimal path as directed by the MPRs to the destination. In this work, the routing flow of the protocol to get rid of other processes that may consume more of the limited network resources hence the use of the AES algorithm was modified. Nodes can exchange updates and route table information. Adaptive routing allows the routing path to change over time as the topology in which the nodes operate in is ever-changing. Another role player component in our framework is packet switching. Packet switching is regarded as a higher-level decision-making entity responsible for driving packets from source to destination.

The routing protocol, OLSR's flow chart in the proposed framework is shown in Figure 3 where we propose a few changes towards some of the traditional operations of the protocol to accommodate QoS and increase efficiency in terms of performance. Figure 3 shows the flow chart of our modified OLSR protocol. PRs still play a critical role in terms of propagating TC messages through to the routing table. An un-authenticated node will be regarded as a malicious node and will be isolated from the network. At first, the node will be sent a fake HELLO message then selected as an MPR then blacklisted at the routing table. OLSR needs constant updating of its route table due to the nature of MANETs and its dynamic topology.

After routing, packets are sent to the data link layer where there are packet queue management, MAC differentiator, and available bandwidth estimator. Under packet queue management, there is packet scheduling, priority scheduling, and packet forwarding.

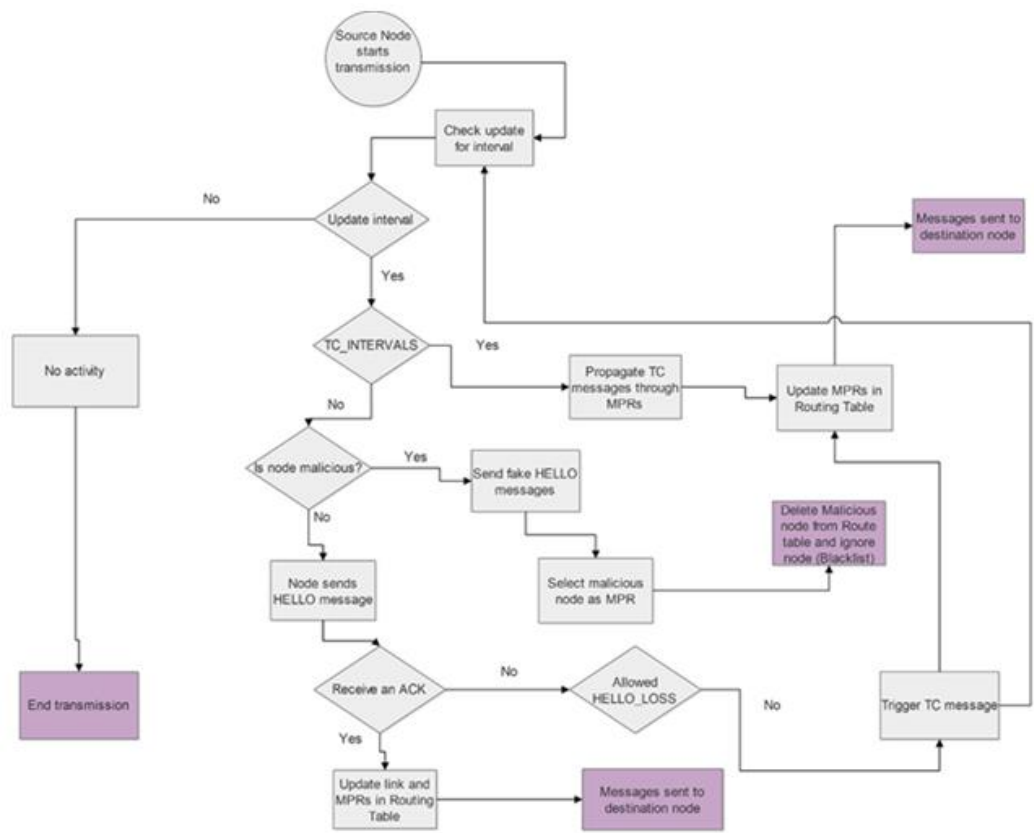

### 7.5.4. Packet Queue Management

When packets arrive at the data link layer they are processed according to the First-In-First-Out rule. The first job to come in is scheduled first. The packet scheduler is responsible for providing the actual memory used for storing packets and providing a firewall used against malicious nodes whose intent is to selfishly use up network resources.

## 8. CONCLUSIONS

In this paper, we have presented and discussed a security framework from the perspective of QoS by using MANET routing protocol, OLSR. The architecture of the QoS-aware security framework was presented and its components were explained with the aim of QoS delivery in video streaming applications. The functions within the framework were explained in accordance to their respective interconnected layers. It is of paramount importance that whatever scheme is used in this architecture, the network resources are spared as best as possible since all nodes are moving randomly and in unfriendly topologies. Most approaches to network security do not consider the aspect of QoS hence our contribution to develop a QoS-centric framework that will not only look into the security aspect but also QoS delivery in the network. Computing the QoS of a MANET is due to the dynamic topology in which the mobile nodes operate in.

A custom flow chart of the routing protocol, OLSR, was also presented as part of our contribution to QoS-aware adaptive routing and improving the existing OLSR routing protocol. As future work, the utmost intention is to evaluate the proposed framework at both the network and application layers. This is a work in progress paper on its final evaluation stages.

## 9. ACKNOWLEDGMENTS

This work would not be possible without the support from the Faculty of Natural and Agricultural Sciences at the North-West University and our partners at TETCOS, India who provided the test-bed for our on-going practical and evaluation of the proposed framework.